\documentclass[5p,twocolumn]{elsarticle}

\usepackage{graphicx}
\usepackage{dcolumn}
\usepackage{pifont}
\usepackage{bm}
\usepackage{multirow}
\usepackage{amsmath}
\usepackage{float}
\usepackage{txfonts}
\usepackage[version=3]{mhchem}

\journal{Chem Commun}

\begin{document}
\title{Structural and optoelectronic properties of the inorganic perovskites \ce{AGeX3} (A = Cs, Rb; X = I, Br, Cl) for solar cell application}

\author[kimuniv-m,kimuniv-n]{Un-Gi Jong}
\author[kimuniv-m]{Chol-Jun Yu\corref{cor}}
\ead{cj.yu@ryongnamsan.edu.kp}
\author[kimuniv-m]{Yun-Hyok Kye}
\author[kimuniv-m]{Yong-Guk Choe}
\author[ntu]{Hao Wei}
\author[ntu]{Shuzhou Li}

\cortext[cor]{Corresponding author}

\address[kimuniv-m]{Chair of Computational Materials Design, Faculty of Materials Science, Kim Il Sung University, \\ Ryongnam-Dong, Taesong District, Pyongyang, Democratic People's Republic of Korea}
\address[kimuniv-n]{Natural Science Centre, Kim Il Sung University, Ryongnam-Dong, Taesong District, Pyongyang, Democratic People's Republic of Korea}
\address[ntu]{School of Materials Science and Engineering, Nanyang Technological University, Nanyang Avenue, 639798, Singapore}

\begin{abstract}
We predict the structural, electronic and optic properties of the inorganic Ge-based halide perovskites \ce{AGeX3} (A = Cs, Rb; X = I, Br, Cl) by using first-principles method. In particular, absolute electronic energy band levels are calculated using two different surface terminations of each compound, reproducing the experimental band alignment.
\end{abstract}


\maketitle

Recently photovoltaic solar cells have been revolutionized by adopting the halide perovskites with a chemical formula of \ce{ABX3} as a photoaborber~\cite{Kojima,WSYang}.
Initiated by the report of Kojima \textit{et al.}~\cite{Kojima} in 2009, the power conversion efficiency (PCE) of the perovskite solar cells (PSCs) has rapidly jumped from the initial value of 3.8\% to over 22.1\%~\cite{WSYang} within less than ten years.
The most widely investigated PSCs are based on the hybrid organic-inorganic iodide \ce{CH3NH3PbI3} (or \ce{MAPbI3}), where organic \ce{MA+} cation, metallic \ce{Pb^{2+}} cation and \ce{I-} anion occupy the A, B and X sites.
However, the presence of toxic element Pb~\cite{Chatterjee18jmca,Feng14jpcc} and the chemical instability originated from the organic moiety MA~\cite{Berhe,Manser,Jong18jmca,kye18} cause serious concerns on environmental issue and large-scale commercialization.

In order to address these problems, compositional tuning has been introduced; substitutions of nontoxic \ce{Sn^{2+}} or \ce{Ge^{2+}} cation for the \ce{Pb^{2+}} cation, and/or inorganic \ce{Cs+} or \ce{Rb+} cation for the organic \ce{MA+} cation~\cite{Chatterjee18jmca,Feng14jpcc,Protesescu,Swarnkar,Kovalsky}.
Accordingly, \ce{CsSnI3} has been synthesized in cubic phase, but the PCE of the cell was only up to 2\%~\cite{Chen12,Kumar14,Sabba15}.
In parallel to these experiments, theoretical investigations have been performed, revealing that such a coarse efficiency is associated with the large mismatch of band alignment between tin-based perovskite photoabsorber and charge carrier conductors~\cite{Jung17cm,RYang17jpcl}.
Moreover, tin-based halide perovskites have a serious problem of the susceptibility of tin oxidation from the +2 to the +4 state upon exposure to ambient air, resulting in much less research effort towards tin-based PSCs. 
In such situation, germanium-based inorganic perovskites \ce{CsGeX3} (X=I, Br, Cl) have emerged as potential alternatives to tin-based perovskites for lead free and long-term PSCs~\cite{Thirumal15jmca,Ming16jmca,Walters18jpcl,Mao18jpcc}.
More importantly, \ce{CsGeX3} does not suffer from the phase transition to the low dimensional yellow phase (nonperovskite structure) at room temperature; it exists in stable three dimensional (3D) perovskite structure with a small rhombohedral distortion in $R3m$ space group up to 350 $^\circ$C~\cite{Stoumpos13,Stoumpos15,Tang05jpcm,Tang09jjap}.
Regarding the A-site, substitution of Rb for Cs can be expected to enhance the film quality and applicability in solar cell devices~\cite{Jiang18jpcc,Jung17cm}.
To the best of our knowledge, however, neither experimental nor theoretical studies on \ce{RbGeX3} could be found although some experimental and theoretical studies on \ce{CsGeX3}~\cite{Thirumal15jmca,Ming16jmca,Walters18jpcl}.
Therefore, it is necessary to get a comprehensive insight into the structural and optoelectronic properties of \ce{RbGeX3} in comparison with \ce{CsGeX3} for solar cell applications.

\begin{table*}[!th]
\scriptsize
\caption{\label{tab_properties}The Goldschmidt's tolerance factor $t_G$, lattice constant ($a$) and angle ($\alpha$), dielectric constant ($\epsilon$), effective masses of electron ($m^*_e$) and hole ($m^*_h$) and reduced mass ($m^*_r$), exciton binding energy ($E_b$), and band gap ($E_g$) in \ce{AGeX3} (A = Cs, Rb; X = I, Br, Cl). The XC functionals are PBEsol and HSE06 without and with spin-orbit coupling (SOC) effect.}
\begin{tabular}{l@{\hspace{3pt}}c@{\hspace{3pt}}c@{\hspace{3pt}}c@{\hspace{3pt}} c@{\hspace{3pt}}c@{\hspace{3pt}}c@{\hspace{3pt}}c@{\hspace{3pt}} c@{\hspace{3pt}}c@{\hspace{3pt}}c@{\hspace{3pt}}c@{\hspace{3pt}} c@{\hspace{3pt}}c@{\hspace{3pt}}c@{\hspace{3pt}}c@{\hspace{3pt}} c@{\hspace{3pt}}c@{\hspace{3pt}}c@{\hspace{3pt}}c@{\hspace{3pt}} c@{\hspace{3pt}}c@{\hspace{3pt}}c@{\hspace{3pt}}c@{\hspace{3pt}}c} 
\hline
 \multicolumn{1}{c}{Compound} & \multicolumn{1}{c}{$t_G$} & & \multicolumn{3}{c}{$a$ (\AA)}  & &  \multicolumn{3}{c}{$\alpha$  ($^\circ$)} & &   \multicolumn{2}{c}{$\epsilon$ }& & $m^*_e$ & $m^*_h$ & $m^*_r$ & $E_b$ (meV) & & \multicolumn{6}{c}{$E_g$ (eV)} \\
\cline{4-6}  \cline{8-10} \cline{12-13} \cline{15-18}  \cline{20-25}
 & & & PBEsol & Exp. & Theo. & & PBEsol & Exp. & Theo. & & PBEsol & Theo. & & \multicolumn{4}{c}{PBEsol}& & PBEsol & PBEsol+SOC & HSE06 & HSE06+SOC & Exp. & Theo. \\
\hline
\ce{CsGeI3} & 0.93 & & 5.94 & 5.98$^a$ & 6.13$^e$ & & 88.68 & 88.60$^a$ & 88.28$^e$ & &  10.40 & 9.98$^g$ & &0.15 & 0.16 & 0.08 & 9.9 & & 1.19 & 1.06 & 1.64 & 1.52 &1.63$^b$ & 1.41$^e$\\
\ce{CsGeBr3} & 0.95 & & 5.56 & 5.64$^b$ & 5.78$^e$ & & 89.15 & 88.79$^b$ & 88.53$^e$ & &  8.05 & 6.52$^g$ & &0.18 & 0.19 & 0.09 & 19.4 & & 1.46 & 1.42 & 2.34 & 1.77 &2.39$^d$ & 1.66$^e$ \\
\ce{CsGeCl3} & 0.97 & & 5.33 & 5.43$^b$ & 5.54$^e$ & & 89.88 & 89.72$^b$ & 88.95$^e$ & &  5.10 & 4.30$^g$ & & 0.27 & 0.28 & 0.14 & 71.9 & & 2.13 & 2.08 & 3.24 & 2.28 &3.40$^d$ & 2.22$^e$ \\
\ce{RbGeI3}  & 0.90 & & 5.91 & $-$ & $-$ & & 87.79 & $-$& $-$ & & 11.55 & $-$ & & 0.16 & 0.17 & 0.08 &  8.5 & & 1.31 & 1.18 & 1.78 & 1.67 & $-$& $-$  \\
\ce{RbGeBr3}& 0.92 & & 5.53 & $-$& $-$ & & 87.99 & $-$& $-$ & & 10.59 & $-$ & &0.18 & 0.18 & 0.09 &  11.1 & & 1.57 & 1.53 & 2.40 & 2.04 & $-$& $-$ \\
\ce{RbGeCl3}& 0.93 & & 5.27 & $-$& $-$ & & 89.66 & $-$& $-$ & &  6.01 & $-$ & & 0.24 & 0.24 & 0.12 & 44.8 & & 2.16 & 2.12 & 3.28 & 2.05 & $-$& $-$ \\
\hline
\end{tabular}
$^a$Ref.~\cite{Tang05jpcm}, $^b$Ref.~\cite{Thirumal15jmca}, $^c$Ref.~\cite{Stoumpos15}, $^d$Ref.~\cite{Lin08om}, $^f$Ref.~\cite{Jong17jps}. \\
$^e$The lattice parameters were determined with PBE functional, while the band gaps with HSE06+SOC~\cite{Walters18jpcl}. \\
 $^g$The dielectric constants were calculated with PBE functional~\cite{Huang16prb}.
\end{table*}

In the present work we perform theoretical investigation based on state-of-the-art density functional theory (DFT) to get a comprehensive understanding of germanium-based halide perovskites \ce{AGeX3} (A = Cs, Rb; X = I, Br, Cl) for solar cell applications (for details of method, see ESI$\dag$). Systematic comparison of structural, electronic and optic properties of these crystalline materials is provided, paying special attention to the band alignments that are dependent on the surface terminations. The results shed light on the systematic variation of the optoelectronic properties and on the feasibility of modulating the energy level alignment upon exchange of A-site cation or X-site anion in \ce{AGeX3}.

The family of inorganic Ge-based halide perovskites \ce{AGeX3} is expected to stabilize in the rhombohedral phase with $R3m$ space group at room temperature, as experimentally identified for \ce{CsGeX3}~\cite{Stoumpos13,Stoumpos15,Tang05jpcm,Tang09jjap}.
As a preliminary test of perovskite structure formability, we first analysed their Goldschmidt's tolerance factors, calculated by $t_G=(r_{\ce{A}}+r_{\ce{X}})/\sqrt{2}(r_{\ce{B}}+r_{\ce{X}})$ using the Shannon ionic radii~\cite{Kieslich}. Based on the established fact that $t_G$ ranging from 0.8 to 1.0 can support formation of perovskite structure, all six compounds are expected to crystallize in a stable perovskite phase due to their proper tolerance factors within $0.90<t_G<0.97$ (Table~\ref{tab_properties}). Through the structural optimizations, it was found that the corner sharing \ce{GeX6} octahedra were slightly distorted because of \ce{Ge^{2+}} cation off-centering as indentified in the cubic \ce{CsSnX3} and \ce{CsPbX3} by first-principles calculations~\cite{RYang17jpcl} (see ESI,$\dag$ Fig. S1). Such octahedral distortion implies a creation of spontaneous electric polarization that can enhance charge carrier separation and allow the photovoltage to exceed the band gap~\cite{Leguy15nc}.

The calculated lattice constants and angles by use of PBEsol exchange-correlation (XC) functional~\cite{PBEsol} within the generalized gradient approximation (GGA) are listed in Table~\ref{tab_properties}.
Our calculated lattice constants of 5.94, 5.56 and 5.33 \AA~for \ce{CsGeX3} with X = I, Br and Cl are in good agreement with the experimental values~\cite{Tang05jpcm,Thirumal15jmca} with a slight underestimation of $\sim$1.8\%, being better than those by PBE functional~\cite{pbe} with an overestimation of $\sim$2.5\%~\cite{Walters18jpcl}.
On the contrary, lattice angles were overestimated with PBEsol in this work, while they were underestimated with PBE~\cite{Walters18jpcl}.
As the ionic radius of X-site halide anion decreases ({\it i.e.}, going from I to Br to Cl), the lattice constant decreases and the lattice angle becomes close to 90$^\circ$, indicating that the crystalline lattice shrinks and changes from the rhombohedral to the cubic phase.
Such trend of crystalline lattice variation is well coincident with the increasing tendency of tolerance factor from X = I to Cl, reminding that for a perfect cubic perovskite structure $t_G=1$.
These can be associated with the strengthening of chemical bond between Ge and X atoms and subsequently the slackening of Ge atom off-centering in \ce{GeX6} octahedra, possibly resulting in an enhancement of chemical stability going from X = I to Cl.
Upon exchange of A-site cation from Cs to Rb, the lattice parameters slightly decrease in accordance with a reduction of ionic radius.

\begin{figure*}[!t]
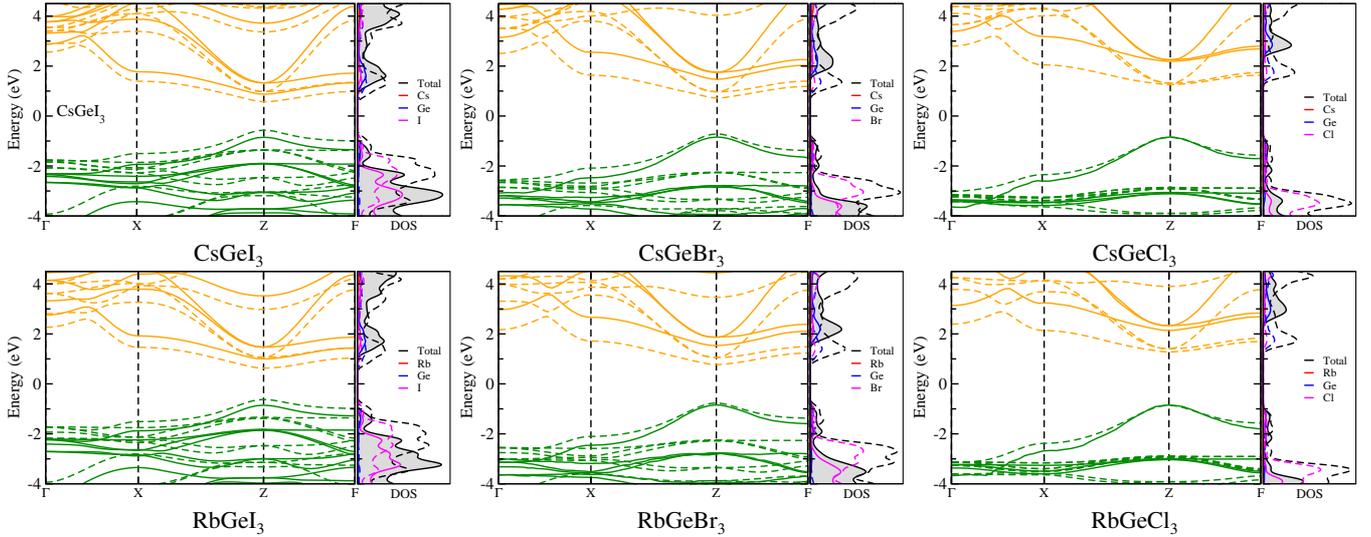

\small
\begin{center}
\begin{tabular}{c@{\hspace{3pt}}c@{\hspace{3pt}}c}
\includegraphics[clip=true,scale=0.29]{fig1a.eps} &
\includegraphics[clip=true,scale=0.29]{fig1b.eps} &
\includegraphics[clip=true,scale=0.29]{fig1c.eps} \\
\ce{CsGeI3} & \ce{CsGeBr3} & \ce{CsGeCl3} \\ 
\includegraphics[clip=true,scale=0.29]{fig1d.eps} &
\includegraphics[clip=true,scale=0.29]{fig1e.eps} &
\includegraphics[clip=true,scale=0.29]{fig1f.eps} \\
\ce{RbGeI3} & \ce{RbGeBr3} & \ce{RbGeCl3}
\end{tabular} 
\end{center}
\caption{\label{fig_bandDOS}The calculated electronic band structures and density of states (DOS) of inorganic halide perovskites \ce{AGeX3} (A = Cs, Rb; X = I, Br, Cl) by use of PBEsol (dotted line) and HSE06 (solid line) functionals. Green and orange colours denote valence and conduction bands, respectively.}
\end{figure*}

In a next step, we calculated their electronic structures by use of different levels of DFT method; PBEsol and hybrid HSE06~\cite{HSE03jcp} XC functionals without or with a consideration of spin-orbit coupling (SOC) effect.
For the hybrid perovskite \ce{MAPbI3}, the GGA functionals were found to yield a band gap coincident with experiment, due to a fortuitous error cancellation between the GGA underestimation and the overestimation by ignoring SOC effect~\cite{Even13,Umari}. However, they failed in reproducing the experimental band gaps for \ce{CsGeX3} (Table~\ref{tab_properties}); significant underestimation with PBEsol indicates the relatively weak SOC effect in the Ge-based perovskites. Such underestimation of band gap becomes more severe for lighter halogen atoms going from I to Cl due to becoming weaker relativistic effect~\cite{Jong17jps} (see ESI,$\dag$ Fig. S2). When including the SOC effect ({\it i.e.}, PBEsol+SOC), band gaps were further underestimated. On the other hand, the hybrid functional HSE06 gave the band gaps in good agreement with the experimental values for \ce{CsGeX3} with absolute accuracies of $+$0.01, $-$0.05 and $-$0.16 eV for X = I, Br and Cl (for \ce{CsPbI3}, GW without SOC effect can also yield a reliable band gap~\cite{jong18prb}). Accordingly, the HSE06+SOC calculations produced the underestimated band gaps in contrast to the cases of \ce{MAPbX3}, for which the hybrid functional or GW method coulpled with SOC effect can give the correct band gaps~\cite{Even13,Umari}.

As shown in Fig.~\ref{fig_bandDOS}, HSE06 calculations properly push up the conduction bands (CBs) while push down the valence bands (VBs) calculated by PBEsol. It was found that transition of electrons from the valence band maximum (VBM) to the conduction band minimum (CBM) can occur in direct way at the edge point Z (0.5, 0.5, 0.5) of the Brillouin zone (BZ). This is similar to the cubic \ce{MAPbX3} (R point)~\cite{Jong17jps}, and is useful to the generation of charge carriers (conduction electrons and holes) by absorbing light. Based on the HSE06 calculations, \ce{CsGeI3} can be said to be suitable for applications of light-absorber due to its proper band gap of 1.64 eV, while \ce{RbGeI3} with a band gap of 1.78 eV can be used as a top cell material in the tandem solar cell system. When exchanging I atom for Br or Cl atom, the band gap becomes larger as over 2.3 or 3.2 eV, implying possible applications of charge carrier conducting materials. Through the analysis of the electronic density of states (DOS), it was found that the VBM is dominated by X $p$ states coupling with Ge $4s$ state, while the CBM is composed of strong antibonding coupling of Ge $4p$ states and X $s$ states (Fig.~\ref{fig_bandDOS}).

In order to attain a deeper understanding of the potentiality of these perovskites to be used as light absorber or charge carrier conductor, we evaluated the various optic properties such as effective masses of electron and hole, exciton binding energies, dielectric constants, and photoabsorption coefficients by use of the PBEsol functional. The effective masses of the charge carriers were obtained by postprocessing the refined band structures around the Z point of BZ, which clearly show parabolic characteristics. Our calculated effective masses of electron and hole for \ce{CsGeI3} (0.15$m_e$ and 0.16$m_e$ in Table~\ref{tab_properties}) are slightly smaller than the previous values (0.21$m_e$ and 0.22$m_e$ in Ref.~\cite{Ming16jmca}), and are comparable with those for \ce{MAPbI3} (0.12 $\sim$ 0.19$m_e$ and 0.15 $\sim$ 0.24$m_e$~\cite{Jong16prb,Jong17jps}), indicating high mobility of charge carriers. As going from I to Cl, the effective masses become larger but still sufficiently small for charge carrier conductors. Upon exchange of Cs for Rb, the values are more or less unchangeable, or even become smaller for the case of X = Cl, giving a prediction of no detriment of optic properties.

Then, using the density functional perturbation theory (DFPT) approach including the effect of atomic displacements, we calculated the frequency dependent macroscopic dielectric functions, from which the static dielectric constants were extracted at the zero photon energy. It should be noted that the Bethe-Salpeter approach considering the excitonic ({\it i.e.}, the electron-hole interaction) and many-body effects gives the almost same dielectric constants to those from DFPT~\cite{Jong17jps}. The static dielectric constant ($\varepsilon$) of \ce{CsGeI3} was determined to be 10.40, which is higher than that of the cubic \ce{MAPbI3} ($\sim$5~\cite{Jong17jps}), indicating more effective screening of electron-hole interaction by lattice dielectric response. As the atomic number of X-site halide anion decreases, $\varepsilon$ decreases to 5.10 for \ce{CsGeI3}, which can be closely associated with a smaller degree of Ge atom off-centering for a smaller ionic radius halide anion (leading to the \ce{GeX6} octahedral distortion) as discussed above. Replacing Cs with Rb causes a slight increase of static dielectric constants in \ce{AGeX3} (X = I, Br, Cl), indicating a possible improvement of charge carrier mobility by such replacement.

As a crucial factor for determining the optic and transport properties, the exciton binding energies ($E_b$) were estimated based on the calculated effective masses and static dielectric constants (see Table~\ref{tab_properties}). For the rhombohedral \ce{CsGeI3}, it was found to be 9.9 meV, which is distinctly smaller than the value of 45 $\sim$ 50 meV for the cubic \ce{MAPbI3}~\cite{Jong16prb,Jong17jps}, indicating that the influence of the \ce{GeX6} octahedral distortion on exciton binding is more pronounced than the screening of MA cation rotational motion. As earlier expected, $E_b$ increases to 19.4 and 71.9 meV for \ce{CsGeBr3} and \ce{CsGeCl3} due to the decrease of the static dielectric constant going from X = I to Br and Cl. As such, it decreases slightly upon the replacement of Cs by Rb. Photoabsorption coefficients as a function of photon energy were calculated using the frequency dependent dielectric functions, showing a gradual shift of the absorption onset and the first peak position and a lowering of the first peak magnitude as the atomic number of X-site atom decreases (see ESI,$\dag$ Fig. S3).

\begin{figure*}[!t]
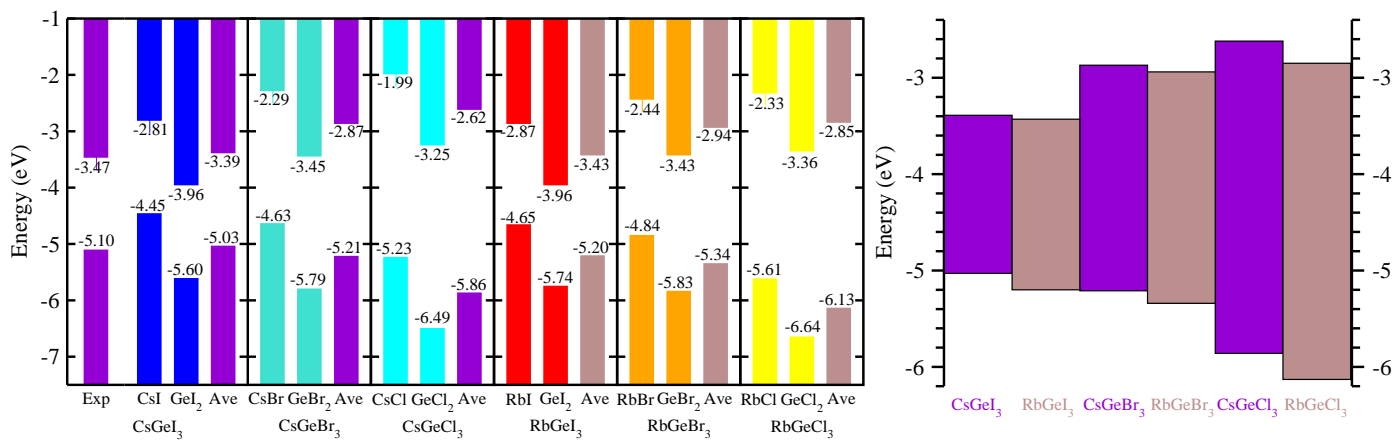

\begin{center}
\includegraphics[clip=true,scale=0.5]{fig2a.eps} 
\includegraphics[clip=true,scale=0.5]{fig2b.eps} 
\end{center}
\caption{\label{fig_alignment}The HSE06-calculated band energy alignment diagrams of \ce{AGeX3} (A = Cs, Rb; X = I, Br, Cl) using AX and \ce{GeX2} terminated (001) surfaces. All energy levels are aligned with respect to the absolute vacuum level, which is set to 0 eV. Exp and Ave mean the experimental and average. Right panel shows the average energy level diagram of \ce{AGeX3} perovskites.}
\end{figure*}
In a final step, we calculated their absolute electronic energy levels of \ce{AGeX3} with respect to an external vacuum level, which are of particluar interest in devising a high-efficiency solar cell system. Since the absolute band energy level depends on the surface termination~\cite{Jung17cm}, we considered two kinds of terminations, AX and \ce{GeX2}, on the (001) surface of each \ce{AGeX3} (see ESI,$\dag$ Fig. S1(b) and (c)). Fig.~\ref{fig_alignment} depicts the VB and CB energy level alignment diagrams of \ce{AGeX3} calculated by use of HSE06 hybrid functional. For the case of \ce{CsGeI3}, we also give the experimental result~\cite{Thirumal15jmca}, which falls properly between our calculated band energy levels of two different surface terminations. For \ce{CsGeX3}, the band energy level differences between the CsX and \ce{GeX2} terminations are about 1.2 eV, while for \ce{RbGeX3} they are around 1.1 eV. Interestingly, the average values of two energy levels of different surface terminations are obtained as $-3.39$ and $-5.03$ eV for \ce{CsGeI3}, which agree well with the experimental values with an absolute accuracy of +0.08 eV. For such a good agreement by averaging, the reason is that the measurment of band energy level, e.g., by photoemission spectroscopy in air~\cite{Thirumal15jmca}, can be done on the surfaces with different indices and terminations. It should be emphasized that the band energy level changes by A-cation replacement of Rb for Cs are much smaller than those by X-anion exchange.

In conclusion, we have investigated the structural and optoelectronic properties of inorganic Ge-based halide perovskites \ce{AGeX3} (A = Cs, Rb; X = I, Br, Cl) with the first-principles calculations by use of PBEsol and HSE06 functionals with and without SOC effect. For the cases of \ce{CsGeX3}, the lattice constants, effective masses of charge carriers, static dielectric constants and exciton binding energies calculated with PBEsol have been confirmed to agree well with experiment. We have shown that averaging the band energy levels of different surface terminations, calculated by use of HSE06 functional, can reproduce the experimental band alignments and band gaps. We predict that replacement of Cs with Rb can offer reasonable flexibility in optoelectronic property matching for solar cell design and optimisation, while X-anion exchange gives rise to big changes, and that these inorganic Ge-based halide perovskites could be candidates for stable and large-scale solar cell development.

This work was supported partially by the State Committee of Science and Technology, Democratic People's Republic of Korea, under the state research project ``Design of Innovative Functional Materials for Energy and Environmental Application'' (No. 2016-20). The calculations have been carried out on the HP Blade System C7000 (HP BL460c) that is owned by Faculty of Materials Science, Kim Il Sung University.

\section*{Conflict of interest}
There are no conflicts to declare.

\bibliographystyle{elsarticle-num-names}
\bibliography{Reference}

\end{document}


\title{Supporting information -- Structural and optoelectronic properties of the inorganic perovskites \ce{AGeX3} (A = Cs, Rb; X = I, Br, Cl) for solar cell application}

\author{Un-Gi Jong$^{a,b}$, Chol-Jun Yu$^a$\footnote{Corresponding author: Chol-Jun Yu, Email: cj.yu@ryongnamsan.edu.kp}, Yun-Hyok Kye$^a$, Yong-Guk Choe$^a$, Hao Wei$^c$, Shuzhou Li$^c$ \\ \\
\small \it $^a$Chair of Computational Materials Design, Faculty of Materials Science, Kim Il Sung University, \\ 
\small \it Ryongnam-Dong, Taesong District, Pyongyang, Democratic People's Republic of Korea \\
\small \it $^b$Natural Science Centre, Kim Il Sung University, Ryongnam-Dong, Taesong District, \\
\small \it Pyongyang, Democratic People's Republic of Korea \\
\small \it $^c$School of Materials Science and Engineering, Nanyang Technological University, \\ 
\small \it Nanyang Avenue, 639798, Singapore}
\date{}
\maketitle

\section*{Computational details}
All calculations in this work were performed by using the pseudopotential plane-wave method as implemented in the Vienna Ab initio Simulation Package (VASP)~\cite{vasp1,vasp2}. Projector augmented wave (PAW)~\cite{paw1,paw2} base functions provided in the package were used to describe the ion-electron interactions. Here, the valence electronic configurations of the atoms are Cs--5s$^2$5p$^6$6s$^1$, Rb--4s$^2$4p$^6$5s$^1$, Cl--3s$^2$3p$^5$, Br--4s$^2$4p$^5$, I--5s$^2$5p$^5$, and Ge--4s$^2$4p$^2$. The Perdew-Burke-Ernzerhof functional for solid (PBEsol)~\cite{PBEsol} within generalized gradient approximation (GGA) was used to describe the exchange-correlation interaction between the valence electrons. As the major computational parameters, the plane-wave cutoff energy was set to 500 eV, and the Monkhorst-Pack special $k$-points were set to (6$\times$6$\times$6) for structural relaxations and (10$\times$10$\times$10) for electronic structure calculations, which can provide an accuracy of total energy as 1 meV per formula unit. The convergence thresholds for total energy and atomic forces were set to 10$^{-6}$ eV and 10$^{-4}$ eV/$\AA$, respectively. To obtain more reliable band gaps, we have also performed electronic structure calculations by use of the HSE06 hybrid functional~\cite{HSE03jcp,HS04jcp,HSE06jcp}. The spin-orbit coupling (SOC) effect has been considered only in the electronic structure calculations. By using density functional perturbation theory (DFPT) method, we calculated the macroscopic frequency-dependent dielectric functions, $\varepsilon$($\omega$)=$\varepsilon_1$($\omega$)+$i\varepsilon_2$($\omega$), and estimated the photo-absorption coefficients $\alpha(\omega)$ as follows~\cite{Jong16prb,Jong17jps},
\begin{equation}
\label{absorption}
\alpha(\omega)=\frac{2\omega}{c}\sqrt{\frac{\sqrt{\varepsilon_1^2(\omega)+\varepsilon_2^2(\omega)}-\varepsilon_1(\omega)}{2}}
\end{equation}
where $c$ and $\omega$ are the light velocity in vacuum and frequency of light wave, respectively. We calculated the effective masses of electron ($m^*_e$) and hole ($m^*_h$) by numerically processing the refined energy band data around Z point. Then, the exciton binding energy ($E_b$) were estimated using the following equation~\cite{Jong16prb,Jong17jps},
\begin{equation}
\label{eig_exciton}
E_b=13.56\frac{m_r^*}{m_e}\frac{1}{\varepsilon_s^2}~(\text{eV})
\end{equation}
where $m_r^*$ ($1/m_r^*=1/m_e^*+m_h^*$) is the reduced effective mass, and $\varepsilon_s$ is the static dielectric constant.

In order to calculate the absolute electronic energy levels, we have built (001) surface slab models with two different terminations, \ce{AX} and \ce{GeX2}, which contain 11 atomic layers and a vacuum layer of 15~$\AA$~thickness (see Fig. S1 (b) and (c)). The outermost atomic positions were fully relaxed, while freezing the innermost three atomic layers corresponding to the bulk positions. For the affordable HSE06 calculations in the surface models, we have adopted the lower plane-wave cutoff energy as 300 eV and smaller $k$-points as (6$\times$6$\times$1). Following the procedures suggested in the previous works~\cite{Jung17cm,Yang15jpcl}, we first calculated the electrostatic potential of (001) relaxed surface slab models, and set an external vacuum level obtained from the electrostatic potential to the reference level. Choosing the Ge $1s$ level as the representative core level, the absolute energy levels such as the valence band maximum (VBM) and conduction band minimum (CBM) of the bulk phases were calculated as follows,
\begin{equation}
\label{equ_vbm}
E_{\text{VBM}} = \epsilon^{\text{KS}}_{\text{VBM}}-(E^{\text{bulk}}_{\text{Ge} 1s}-E^{\text{surf}}_{\text{Ge} 1s}) - V_{\text{vac}}
\end{equation} 
\begin{equation}
\label{equ_cbm}
E_{\text{CBM}} = \epsilon^{\text{KS}}_{\text{CBM}}-(E^{\text{bulk}}_{\text{Ge} 1s}-E^{\text{surf}}_{\text{Ge} 1s})-V_{\text{vac}}
\end{equation} 
where $\epsilon^{\text{KS}}_{\text{VBM}}$ and $\epsilon^{\text{KS}}_{\text{CBM}}$ are the bulk eigenvalues corresponding to the VBM and CBM, $E^{\text{bulk}}_{\text{Ge} 1s}$ and $E^{\text{surf}}_{\text{Ge} 1s}$ are the average Ge 1s levels of the Ge atoms in the bulk phase and in the innermost three layers of the surface slab model, and $V_{\text{vac}}$ is the vacuum level derived from the planar average electrostatic potential of the surface slab models. In addition, exact Hartree-Fock exchange replaced 20~\% of the exchange potential of the Perdew-Burke-Ernzerhof (PBE)~\cite{pbe} functional in all the HSE calculations, providing band gaps in good agreement with experiment. The SOC effect was not considered in these calculations.

\begin{figure}[b]
\begin{center}
\begin{tabular}{lll}
(a) & (b) & (c)\\
\includegraphics[clip=true,scale=0.2]{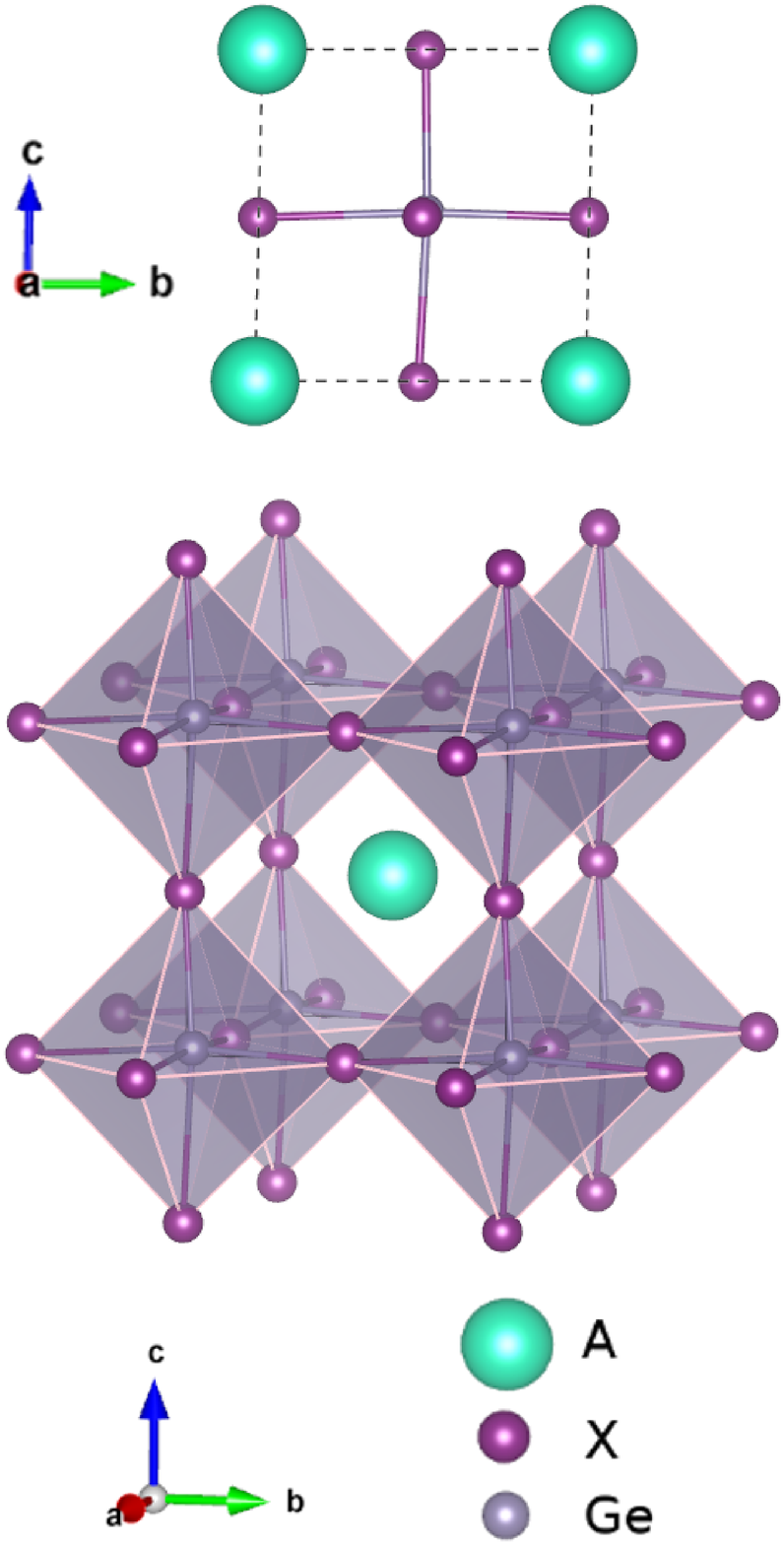} & 
\includegraphics[clip=true,scale=0.2]{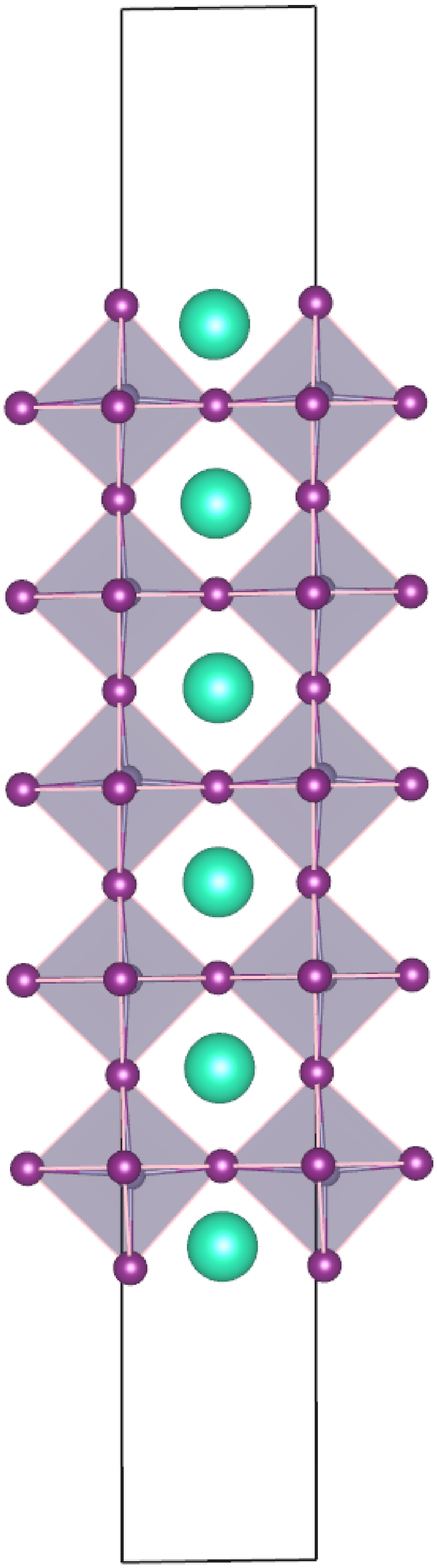} &
\includegraphics[clip=true,scale=0.2]{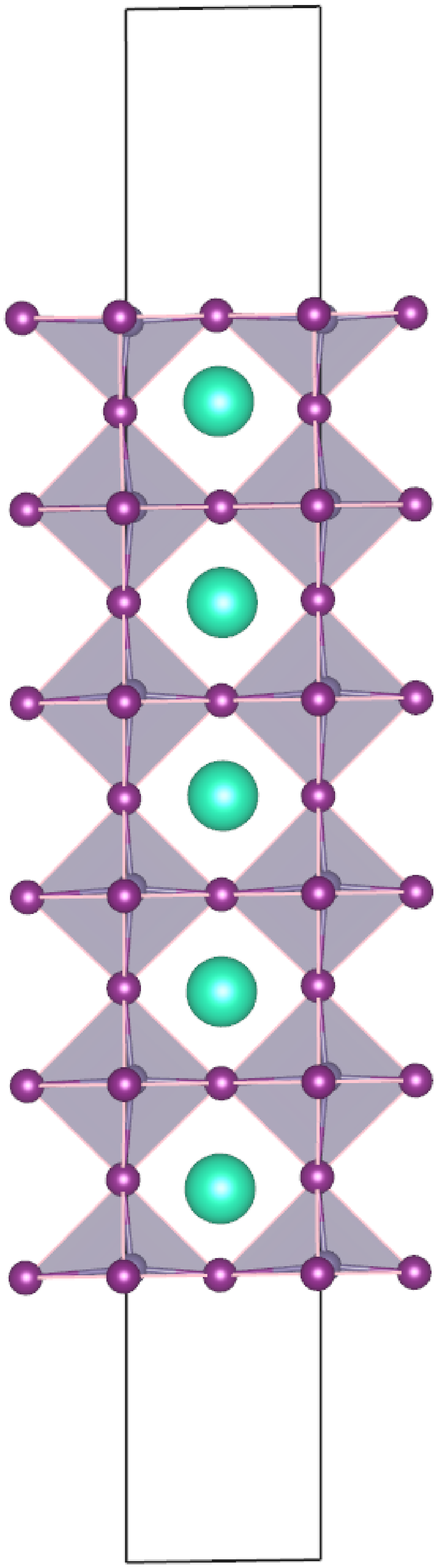} \\
\end{tabular}
\end{center}
Fig. S1 Polyhedral view of (a) bulk and surfaces with two different terminations of (b) AX and (c) \ce{GeX2} of inorganic Ge-based halide perovskites \ce{AGeX3} (A = Cs, Rb; X = I, Br, Cl).
\end{figure}
\begin{figure}[b]
\begin{center}
\includegraphics[clip=true,scale=0.5]{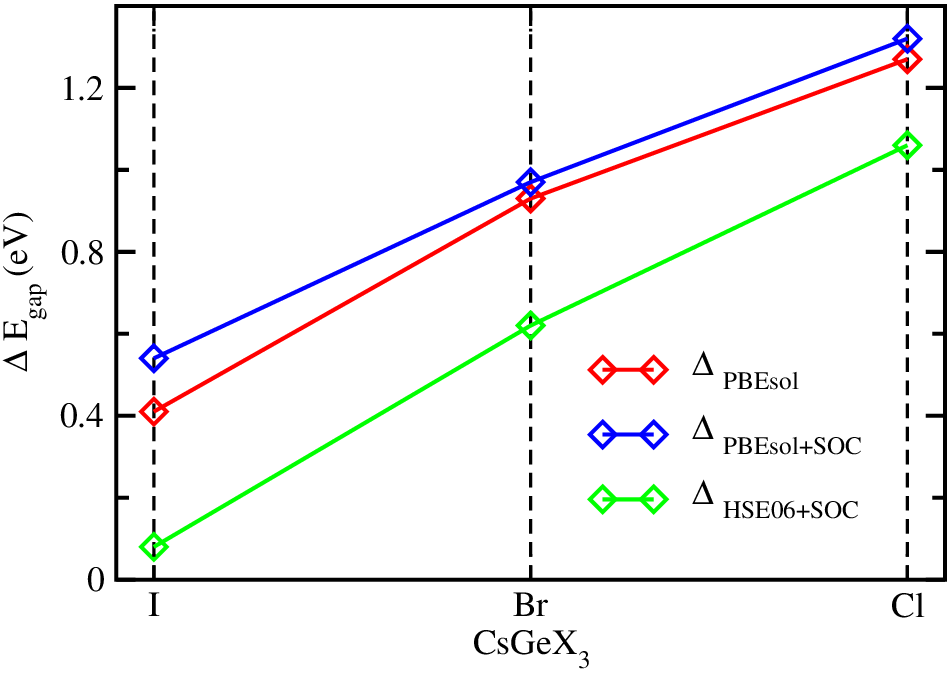} 
\vspace{10pt}
\end{center}
Fig. S2 The calculated band gap deviations from experiment, using $\Delta_{\text{XC}} = E_{\text{gap}}(\text{Exp}) - E_{\text{gap}}(\text{XC})$, where $E_{\text{gap}}(\text{Exp})$ and $E_{\text{gap}}(\text{XC})$ are the experimental and calculated band gaps with XC functionals, respectively.
\end{figure}
\begin{figure}[b]
\begin{center}
\includegraphics[clip=true,scale=0.5]{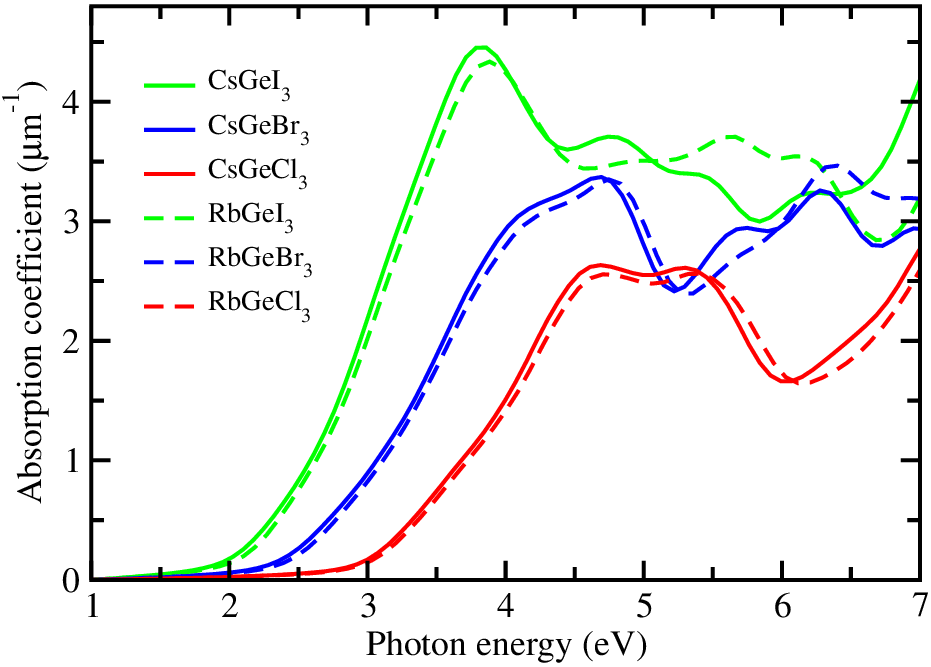} 
\vspace{10pt}
\end{center}
Fig. S3 The calculated photo-absorption coefficients by using the DFPT method with the PBEsol functional.
\end{figure}

\bibliographystyle{unsrt}
\bibliography{Reference}